\documentclass[5p]{elsarticle}

\usepackage[utf8]{inputenc}
\usepackage{color}

\usepackage[colorlinks=true,allcolors=blue,pdfauthor=author]{hyperref}
\biboptions{sort&compress}

\usepackage{amsbsy}
\usepackage{amsfonts}
\usepackage{amsmath}
\usepackage{amssymb}
\usepackage{amsthm}
\usepackage{latexsym}
\usepackage{mathtools}

\usepackage{graphicx}
\usepackage{epsfig}
\usepackage{epstopdf}
\usepackage{svg}
\usepackage[caption=false]{subfig}
\usepackage[font=normal,labelfont=bf]{caption}

\usepackage{booktabs}
\usepackage{tabularx}

\providecommand{\nvec}[1]{\hat{\boldsymbol{#1}}}

\providecommand{\about}[0]{\raise.17ex\hbox{$\scriptstyle\sim$}}

\definecolor{darkred}{rgb}{0.8, 0.0, 0.0}
\definecolor{darkgreen}{rgb}{0.0, 0.8, 0.0}
\definecolor{darkblue}{rgb}{0.0, 0.0, 0.8}

\providecommand{\vanish}[1]{}

\graphicspath{ {./figures/} }

\begin{document}

\begin{frontmatter}

\title{Simulated surface diffusion in nanoporous gold and its dependence on surface curvature}

\author[UCD_MS]{Conner Marie Winkeljohn}
\ead{cmwinkeljohn@ucdavis.edu}

\author[UCD_MS]{Sadi Md Shahriar}

\author[UCD_EE]{Erkin Seker}

\author[UCD_MS]{Jeremy K. Mason\corref{cor}}
\ead{jkmason@ucdavis.edu}

\address[UCD_MS]{Department of Materials Science and Engineering, University of California, Davis, Davis, CA 95616, USA}

\address[UCD_EE]{Department of Electrical and Computer Engineering, University of California, Davis, Davis, CA 95616, USA}

\cortext[cor]{Corresponding author}

\begin{abstract}
The morphological evolution of nanoporous gold is generally believed to be governed by surface diffusion.
This work specifically explores the dependence of mass transport by surface diffusion on the curvature of a gold surface.
The surface diffusivity is estimated by molecular dynamics simulations for a variety of surfaces of constant mean curvature, eliminating any chemical potential gradients and allowing the possible dependence of the surface diffusivity on mean curvature to be isolated.
The apparent surface diffusivity is found to have an activation energy of $\about 0.74$ eV with a weak dependence on curvature, but is consistent with the values reported in the literature.
The apparent concentration of mobile surface atoms is found to be highly variable, having an Arrhenius dependence on temperature with an activation energy that also has a weak curvature dependence.
These activation energies depend on curvature in such a way that the rate of mass transport by surface diffusion is nearly independent of curvature, but with a higher activation energy of $\about 1.01$ eV.
The curvature dependencies of the apparent surface diffusivity and concentration of mobile surface atoms is believed to be related to the expected lifetime of a mobile surface atom, and has the practical consequence that a simulation study that does not account for this finite lifetime could underestimate the activation energy for mass transport via surface diffusion by $\about 0.27$ eV.
\end{abstract}


\end{frontmatter}


\section{Introduction}
\label{sec:introduction}

Nanoporous gold (np-Au) has a bicontinuous structure with interconnected pores and ligaments that have mean widths from nanometers to microns~\cite{McCue2018}.
The interface between the ligaments and pores create a large surface area, and given the characteristic length scale, large surface curvatures as well.
These unique features give rise to many of the desirable properties of np-Au, resulting in its use in a variety of applications including catalysis~\cite{Wittstock2010,Biener2015,Wang2021}, optical sensors~\cite{Qi2014,Zhang2013,Ruffino2020}, biomedical devices~\cite{Chapman2015,Daggumati2016,cseker2018nanoporous}, and battery electrodes~\cite{Yu2011,Wang2021,Gonalves2021}.
One reason for this diversity of uses is that many of its notable properties (e.g., optical properties and mechanical strength) are strongly dependent on a microstructure that can be tailored to each application.
Specifically, the means and standard deviations of the distributions of pore and ligament widths, and the anisotropy of all four of these quantities, can be used to modify the properties of np-Au.
The implication is that precisely controlling the morphology and topology of np-Au is of considerable practical interest.

np-Au is produced by chemically or electrochemically dealloying a silver-gold alloy.
There has already been extensive research on controlling the np-Au morphology by means of the fabrication parameters, including the choice of etching solution and the initial precursor alloy constituents~\cite{Erlebacher2001,Snyder2008,Weissmller2009,Erlebacher2009,Wittstock2010}.
While the pore-to-ligament size ratio is an essential microstructural feature, it is highly sensitive to these parameters and therefore difficult to predict~\cite{Fujita2008}.
This has encouraged the community to develop ways to adjust the pore and ligament sizes after fabrication by means of thermally-driven coarsening, which is the most commonly used technique for this purpose.

The vast majority of studies quantify np-Au microstructure by means of SEM images~\cite{Ziehmer2016,Seker2007,McCue2018,Turner2016} from which the width and length distributions of the ligaments and pores can be calculated.
One method of directly calculating ligament widths and lengths involves reducing the structure pixel by pixel to a skeletal system, counting the number of deleted pixels between the surface and the skeleton to determine ligament radius, and then measuring the length of the skeleton branches~\cite{Fujita2008,Gao2023}.
An alternative is to take the length of the thinnest part of each ligament as the ligament width~\cite{Wang2021}.
The literature on this subject is complicated not only by this diversity of definitions, but by the infrequent reporting of standard deviations of ligament width distributions~\cite{McCue2018}.

A small number of studies use 3D imaging techniques to quantify other aspects of the np-Au microstructure including the pore and ligament topology and the distribution of surface curvatures.
The surface curvature at any point can be described by the two principal curvatures $\kappa_1$ and $\kappa_2$ which are defined as the inverse radii of curvature along the two principal directions, with $\kappa_1$ as the principal curvature of greatest magnitude.
Chen et al.~\cite{Chen2010} used X-ray tomography to image structures in 3D and generate a surface mesh, allowing them to find the surface area per volume $S_A$ and the principal curvatures at each surface point.
The probability distribution of finding a pair of principal curvatures at a surface point is called the interface shape distribution (ISD) and is perhaps the standard way to analyze such curvature information~\cite{Chen2010,ChenWiegart2012,Ziehmer2016}.
The ISD appears to be correlated with other geometric quantities, with $S_A$ generally increasing and the volume fraction of pores generally decreasing with the surface mean curvature~\cite{Chen2010,Fujita2008}.
Part of the reason such a wide variety of characterization measurements are used is that the key characteristics which influence morphology evolution have yet to be precisely identified and related to easily measurable features.

The kinetics of thermally-driven coarsening are usually measured by analyzing SEM images before and after thermal annealing.
The consensus in the literature is that this coarsening occurs by surface diffusion, with prior investigations having ruled out bulk diffusion as well as evaporation/condensation mechanisms \cite{Erlebacher2004,Erlebacher2009,Chen2010,Li2019,Gao2023}.
The mass transport is driven by a chemical potential gradient that results from variations in the surface curvature as given by the Gibbs--Thompson equation;
for a surface of mean curvature $H = \kappa_1 + \kappa_2$, the chemical potential $\mu(H)$ of an atom on the curved surface can be related to that of an atom on a flat surface $\mu(0)$ by
\begin{equation}
\label{equ:GibbsThops}
\mu(H)= \mu(0) + \gamma \Omega H
\end{equation}
where $\gamma$ is the surface energy per area and $\Omega$ is the atomic volume.
%
%
Kolluri and Demkowicz~\cite{Kolluri2011} instead suggest that coarsening could occur by means of network restructuring where localized plasticity allows ligaments to collapse onto their neighbors, making the rate of coarsening dependent on the relative density.
However, they considered a structure with grain sizes comparable to the ligament width, whereas others have observed much larger grain sizes in np-Au~\cite{Weissmller2009}.

There is a frequently-investigated hypothesis that thermal coarsening of np-Au leads to statistically self-similar microstructures~\cite{Hakamada2009,Kolluri2011,ChenWiegart2012,Chen2015,Gao2023}.
Given that surface diffusion is the primary mass transport mechanism, a scaling argument suggests that every characteristic length scale $\lambda$ of a statistically self-similar microstructure increases with time $t$ as:
\begin{equation}
\lambda \propto \big(t D_S\big)^n
\label{eq:scaling_relationship}
\end{equation}
where $n$ is the coarsening exponent and $D_S$ is the surface diffusivity~\cite{Herring1950,Gao2023}. 
Measuring $D_S = D_0 \exp[-E_a / (k_B T)]$ as a function of temperature would then allow the activation energy $E_a$ for coarsening to be found.
%
%
If this relationship holds, the characteristic length scale of np-Au could be easily and predictably controlled by changing the temperature or duration of a thermal anneal.

Efforts in the literature along these lines have yielded conflicting results though, with no consensus that thermal coarsening is self similar and a wide variety of reported activation energies.
Early kinetic Monte Carlo (kMC) simulations found that self-similar evolution was unlikely ($E_a = 0.9$ eV) \cite{Erlebacher2011}, but recent kMC simulations concluded that coarsening was self similar ($E_a = 0.65$ eV) based on an analysis of ligament spacing \cite{Li2019}.
An early experimental study found that the initial coarsening was not self-similar ($E_a = 0.643$ eV), \cite{ChenWiegart2012}, but Jeon et al.\ concluded that coarsening was self similar ($E_a = 0.353$ eV) based on 3D reconstructions of np-Au microstructures \cite{Jeon2017}.
Whereas the above studies used coarsening exponents of $n = 4$, McCue et al.\ reported self-similar coarsening with $n = 8$ based on an analysis of published 2D images \cite{McCue2018}, and Gao et al.\ instead reported coarsening that was not self similar with $n = 3.5$ ~\cite{Gao2023}.
Finally, a study by Ziehmer et al.\ using 3D reconstructions to analyze both principal curvatures was inconclusive as to whether the coarsening of np-Au could be described as self-similar, stating that different regimes could characterize early and late stage coarsening~\cite{Ziehmer2016}.

Our conclusion is that thermal coarsening is still likely to be governed by surface diffusion, but that the factors influencing surface diffusion are not sufficiently well understood for models to accurately predict experimental coarsening rates.
For example, if the surface diffusivity depends on the surface curvature, then $D_S$ in Eq.\ \ref{eq:scaling_relationship} would be a function of $\lambda$.
Equation \ref{eq:scaling_relationship} whould then not be expected to hold for any fixed value of $D_S$ during an interval where $\lambda$ changes appreciably, perhaps leading to the conclusion that thermal coarsening was not self-similar.
Such a possibility would be most likely to occur when $\lambda$ is small since it is for those values that the surface curvature most rapidly changes with respect to $\lambda$.
Many of the conflicting results above could then be explained as consequences of differing initial ligament width distributions.

Our intention is to investigate whether the surface diffusivity could in fact depend on the mean curvature of the surface, at least for the ligaments and pores in np-Au.
While there have been several kMC studies considering the coarsening of np-Au~\cite{Erlebacher2011,Li2019}, these techniques require that the relevant atomic migration rates be specified beforehand.
We instead use molecular dynamics (MD) to study surface diffusion while making minimal prior assumptions about the relevant atomic population and migration mechanisms.
Section \ref{sec:methodology} describes both the setup and conduct of the MD simulations and our approach to extracting surface diffusion coefficients.
The results of this analysis are discussed in Sec.\ \ref{sec:resultsanddisc}, where the surface diffusivity and concentration of mobile surface atoms appear to have a slight dependence on the surface mean curvature.
This is believed to be related to the mobile surface atoms having a finite curvature-dependent lifetime \cite{antczak2007jump} that is not accounted for by the analysis, with the true concentration of mobile surface atoms being constant and independent of curvature.
A finite mobile surface atom lifetime would increase the apparent surface atom population (the population of atoms that diffuse on the surface at any point during the simulation) and decrease the apparent surface diffusivity (a mobile surface atom migrates for only a fraction of the simulation time) compared to the true values of these quantities.
The lifetime is expected to decrease with an increasing rate of exchange between the mobile surface atom population and a second population of comparatively immobile atoms, e.g., ones in the bulk.
A curvature-independent rate of mass transport by surface diffusion then implies that the true surface diffusivity is independent of curvature as well.
This raises the general concern that an analysis that does not account for the finite curvature-dependent lifetimes of the mobile surface atoms could find an activation energy for mass transport that differs by $\about 0.27$ eV from the values reported here.
The consequences of this for predicting experimental np-Au coarsening rates are discussed with the conclusions in Sec.\ \ref{sec:conclusion}.

\section{Methods}
\label{sec:methodology}

The purpose of this work is to characterize diffusion on the surface of the ligaments in np-Au, with the intention of eventually using this knowledge to develop predictive models for thermal coarsening and general morphology evolution.
There are several reasons why measuring surface diffusivities by means of MD simulations is difficult though.
First, the requirement for numerical stability limits the time step to about a femtosecond and the overall length of an MD simulation to a few hundreds of nanoseconds, whereas diffusion experimentally occurs on the order of seconds.
While surface diffusion can be accelerated by increasing the temperature, this can be done only up to the melting point of the material without changing the mass transport mechanism.
Second, diffusion is by definition a continuum-level phenomenon, involving atomic concentrations and chemical potentials that are not easily defined at the atomic scale.
Measuring surface diffusion rates in MD therefore requires a sufficient number of atomic migration events to compare the resulting probability distribution of displacements to that predicted by the phenomenological diffusion equations.
This is difficult to achieve in simulations that are often limited by the available computational resources to fewer than a million atoms.
Third, mass transport can be driven by gradients in any one of a variety of intensive thermodynamic quantities, not just by a chemical potential gradient.
If such effects are not accounted for and controlled in the simulation design, the analysis of the MD results could be subject to systematic errors that would confound subsequent modeling efforts.

\subsection{Molecular dynamics simulations}
\label{sec:simulations}

\begin{figure}
    \centering
    \includegraphics[width=0.98\columnwidth]{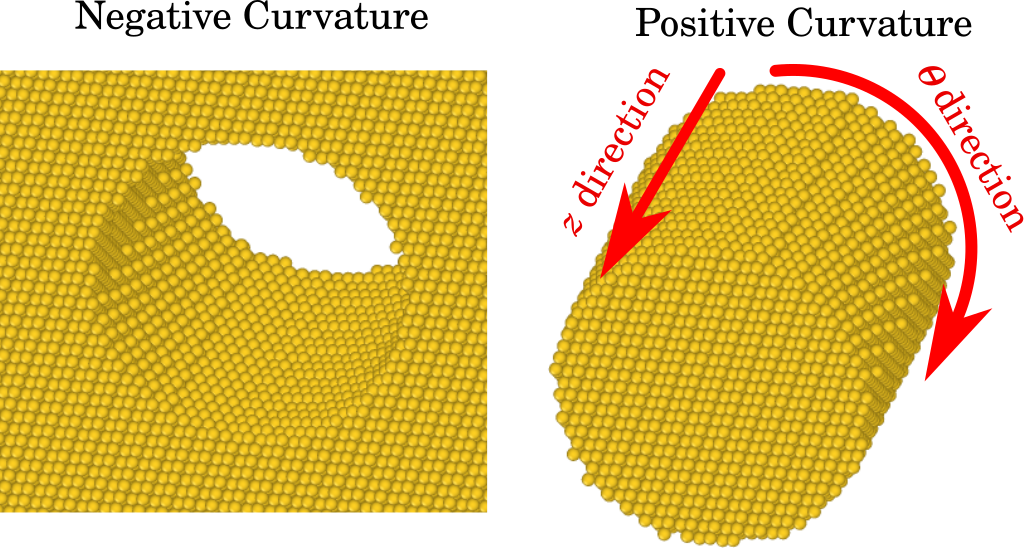}
    \caption{\label{fig:geometry}The simulation cell geometries used to extract surface diffusivity on concave ($\kappa_1<0$) and convex ($\kappa_1>0$) surfaces. $\kappa_2$ is along $\nvec{z}$ and $\kappa_1$ along $\nvec{\theta}$.
    The flat plate used as a zero mean curvature surface is not shown.}
\end{figure}

With a view to the considerations above, this study uses three simulation cell geometries with positive, zero, and negative surface curvatures for our MD simulations of surface diffusion.
Since the simulated atomic displacements will eventually be compared with the predictions of the diffusion equation, it is desirable for the analytical predictions to be as simple as possible.
In the absence of drift, diffusion theory predicts that the distribution of atomic displacements should be identical to those for a random walk.
Hence, it is desirable that the diffusion potential should be constant everywhere, and this effectively requires that the surface be one of constant mean curvature.
Given the characteristic ligaments and pores of the physical system, this suggests that the natural simulation cell geometries are cylinders of positive or negative curvature with $\kappa_2 = 0$ and $\kappa_1 \neq 0$ as in Fig.~\ref{fig:geometry}.
The available computational resources limited the maximum size of the simulation cell and therefore the cylinder radius, resulting in comparatively high mean curvature magnitudes except for one simulation cell containing a flat slab of Au for which the mean curvature was zero.
The five simulation cell geometries that were considered included punch holes of radius $30\ \text{\AA}$ and $34\ \text{\AA}$, a flat slab $35\ \text{\AA}$ thick, and cylinders of radius $34\ \text{\AA}$ and $30\ \text{\AA}$, roughly corresponding to the lower limit of feature sizes experimentally observed in np-Au.
These had corresponding nonzero principal curvatures of $-0.033$, $-0.029$, $0.0$, $0.029$, and $0.033\ \text{\AA}^{-1}$.

The $z$-axis for the simulation box was always parallel to the axis of the punch holes and cylinders.
This had the advantage that the directions of principal curvature in all simulations were aligned with the $\nvec{z}$ and $\nvec{\theta}$ directions, and that the distribution of atomic displacements could be decomposed as a product of independent distributions along these two directions.
The diffusivity is calculated in both the $\nvec{z}$ and $\nvec{\theta}$ directions from these respective displacement distribution data sets for each curvature.

\begin{figure}
    \centering
    \includegraphics[width=0.98\columnwidth]{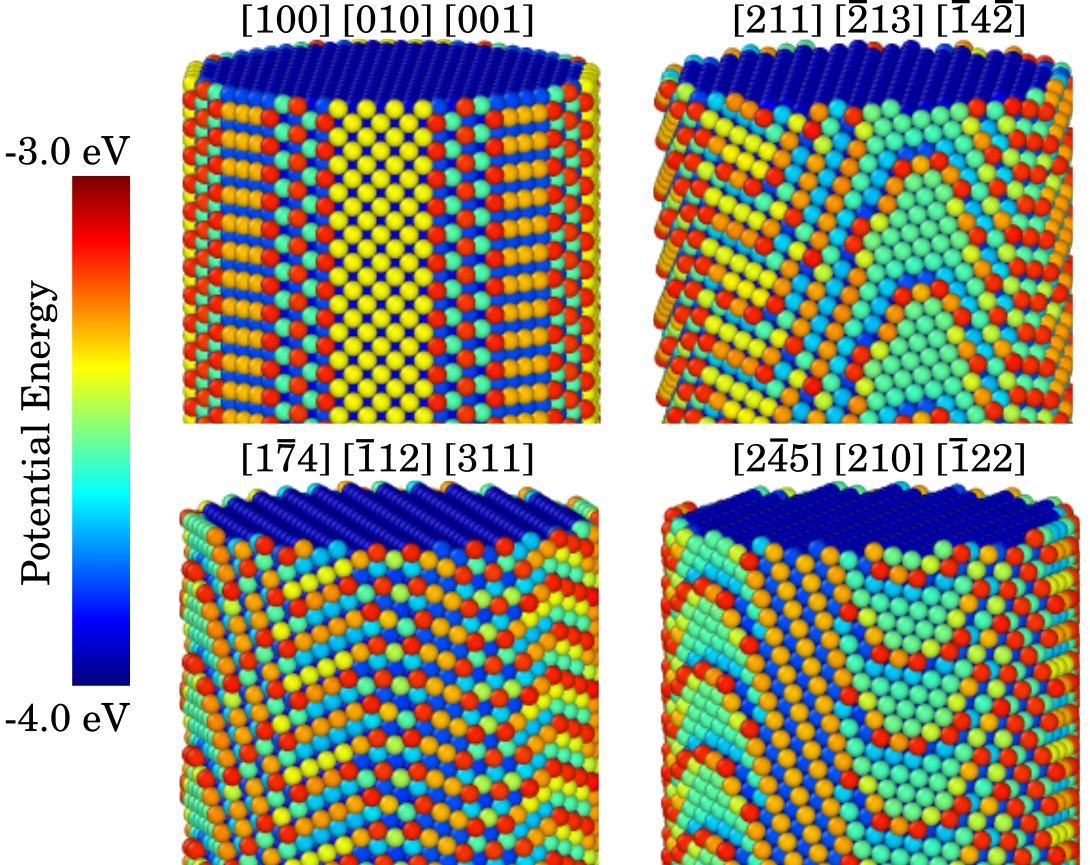}
    \caption{\label{fig:orientations}
    The four different crystal orientations used in our simulations.
    The respective crystallographic directions were oriented along the $x$-, $y$-, and $z$-axes of the simulation cell (the cylinder axis was always along the $z$-axis).
    For example, the [$211$] [$\bar{1}42$] [$\bar{2}13$] orientation aligns the crystallographic direction [$211$] with the $x$-axis of the simulation cell, [$\bar{1}42$] with the $y$-axis, and [$\bar{2}13$] with the $z$-axis.}
\end{figure}

The diffusivity is modeled as being independent of crystallographic orientation given that such a quantity is more likely to be relevant to the coarsening of np-Au when averaged over crystal orientations.
Of course, the crystallographic orientation and the coordination of surface sites does have a significant impact on the surface diffusivity that needs to be accounted for~\cite{Baletto2000,Liu1991}.
Simulating just a single crystallographic orientation would fix the surface facets and potentially bias the coordination distribution's dependence on curvature and the apparent surface diffusivity.
That is, simply varying the radius of a pillar with a fixed crystallographic orientation would result in behavior that depends on the specific choice of orientation and the facet pattern rather than on $\kappa_1$ alone.
Figure \ref{fig:orientations} shows that cylinders with different crystallographic orientations can have dramatically different patterns of surface facets and ridges, with the potential energy of the atom and therefore the activation energy for atomic jumping depending strongly on the type of site.
Moreover, the size of atomic jumps along the $\nvec{z}$ and $\nvec{\theta}$ directions depends explicitly on the corresponding crystallographic directions, causing the distribution of atomic displacements to be discrete rather than continuous as would be predicted by continuum diffusion theory.
Establishing the effective surface diffusivity therefore requires averaging over several crystal orientations.

The periodic boundary conditions require that the simulation cell edge lengths be integer multiples of the crystal lattice periodicity, effectively constraining the crystallographic directions along the coordinate axes to those with single-digit Miller indices.
The four crystallographic orientations shown in Fig.\ \ref{fig:orientations} were used for the cylinders, punch holes, and slabs, with the listed crystallographic directions aligned with the $x$-, $y$-, and $z$-axes of the simulation cell.
These orientations were selected partly to give a variety of angles between the $\langle 111 \rangle$ directions and the $z$-axis, since diffusive jumps are expected to be easiest on $\{ 111 \}$ planes.
The displacement distributions of all four crystallographic orientations were combined into a single data set for a given $\kappa_1$ to find the orientation-averaged diffusivity in the subsequent analysis. 
This had the added advantage of increasing the size of the sampled data to help mitigate random errors.
A comparison between the displacements from a single simulation and those compiled from four simulations, each with a different crystal orientation, is show in Fig.\ \ref{fig:gaussian_smooth}.

All simulations were run using a modified embedded atom method (MEAM) potential in LAMMPS~\cite{LAMMPS}.
The MEAM potential was chosen over more commonly-used EAM potentials for several reasons.
As our study is concerned with the behavior of surface atoms, an interatomic potential with inaccurate surface energies could lead to surface reconstruction and a dramatic change to the surface diffusivity.
While standard EAM potentials \cite{Grochola2005} show good agreement to bulk experimental properties, they are notorious for their poor agreement with experimental surface energies (though recently there has been improvement \cite{becker_trautt_hale_2011}).
Whereas EAM potentials assume that the electron density $\rho$ is spherically symmetric, the MEAM potential is more flexible and allows $\rho$ to depend on angular coordinates~\cite{Lee2003}.
Since our simulations are not only concerned with but dominated by the surface properties of Au rather than the bulk properties, this angular dependence could be crucial for accurately modeling the intended behavior.
Numerical results supporting the selection of the MEAM potential are discussed further in Sec.\ \ref{sec:resultsanddisc}.

All simulations used periodic boundary conditions in all directions and were initialized to be roughly $9$-$10$ nm in the $z$ direction.
The punch hole geometry simulations had $x$ and $y$ lengths of roughly $16$-$20$ nm and contained $106,000$ to $160,000$ atoms, with the larger simulation size required to prevent the periodic images of the punch holes from interacting elastically.
By comparison the cylinder geometry simulations contained only $15,000$ to $20,000$ atoms and the slab geometry simulations contained $19,000$ to $22,000$ atoms, and both of these had $x$ and $y$ lengths of roughly $9$ nm.
After being initialized, all simulations were relaxed by a potential energy minimization in LAMMPS.
This was followed by equilibration for $1$ ns in the isothermal-isobaric ensemble at zero pressure and at the same temperature subsequently used in the production runs to allow for surface reconstruction.
All simulations were finally run for $10$ ns with NPT integration using the Nos{\'e}--Hoover thermostat and barostat, with the atomic displacement data over this interval used for the analysis reported in Sec.~\ref{sec:resultsanddisc}.



\subsection{Fitting diffusion coefficients}
\label{sec:diffusion}

Surfaces are well known to be short circuit diffusion pathways, and the diffusivity of atoms at the surface is expected to be much higher than the diffusivity of atoms in the bulk.
Estimating the diffusivity of the surface atoms therefore requires a way to distinguish between these two populations, ideally while introducing as little bias as possible.
For example, it is not sufficient to simply identify all of the atoms with low coordination numbers at the beginning of the simulation as surface atoms since some of these could eventually become bulk atoms, after which point their behavior should not be included in the statistics.
Nor is it sufficient to include atoms in the statistics only during periods for which their coordination numbers are below an arbitrarily specified cutoff, since the experimental surface diffusivity describes the average behavior of atoms with a variety of environments.

Rather than first identifying the population of surface atoms and then analyzing the behavior of this population, the approach developed here is to do the opposite.
The motion of an individual atom in the absence of a diffusion potential gradient is reasonably modeled as that of a random walker whose one-dimensional displacement $x$ after a time $t$ is distributed as
\begin{equation}
\label{prob_distr}
P(x,t) = \frac{1}{2\sqrt{\pi Dt}}\exp\left(-\frac{x^2}{4Dt}\right)
\end{equation}
where $D$ is the temperature-dependent diffusivity.
Assume that the sets of atoms belonging to the bulk and surface populations are unknown but fixed (this will be reconsidered below), let $D_S$ and $D_B$ respectively be the diffusivities of the surface and bulk populations, and let $n_S$ be the unknown fraction of atoms that are included in the surface population.
Then the probability distribution for the one-dimensional displacement $x$ of an atom selected uniformly at random from all atoms in the simulation cell is
\begin{align}
\label{equ:prob_distr_data}
P(x,t) &= \frac{n_S}{2\sqrt{\pi D_St}}\exp\left(-\frac{x^2}{4D_St}\right) \nonumber \\
&\quad +\frac{1-n_S}{2\sqrt{\pi D_Bt}}\exp\left(-\frac{x^2}{4D_Bt}\right).
\end{align}
Observe that the quantities $D_S$, $D_B$, and $n_S$ can be determined by fitting this equation to the distribution of all atomic displacements without any of the complications described above.
Indeed, since diffusion is a continuum-level phenomenon, one expects to be able to measure surface diffusivities without needing to specify which atoms during which periods of time are considered as belonging to the surface population.

\begin{figure}[t]
    \centering
    \includegraphics[width=\columnwidth]{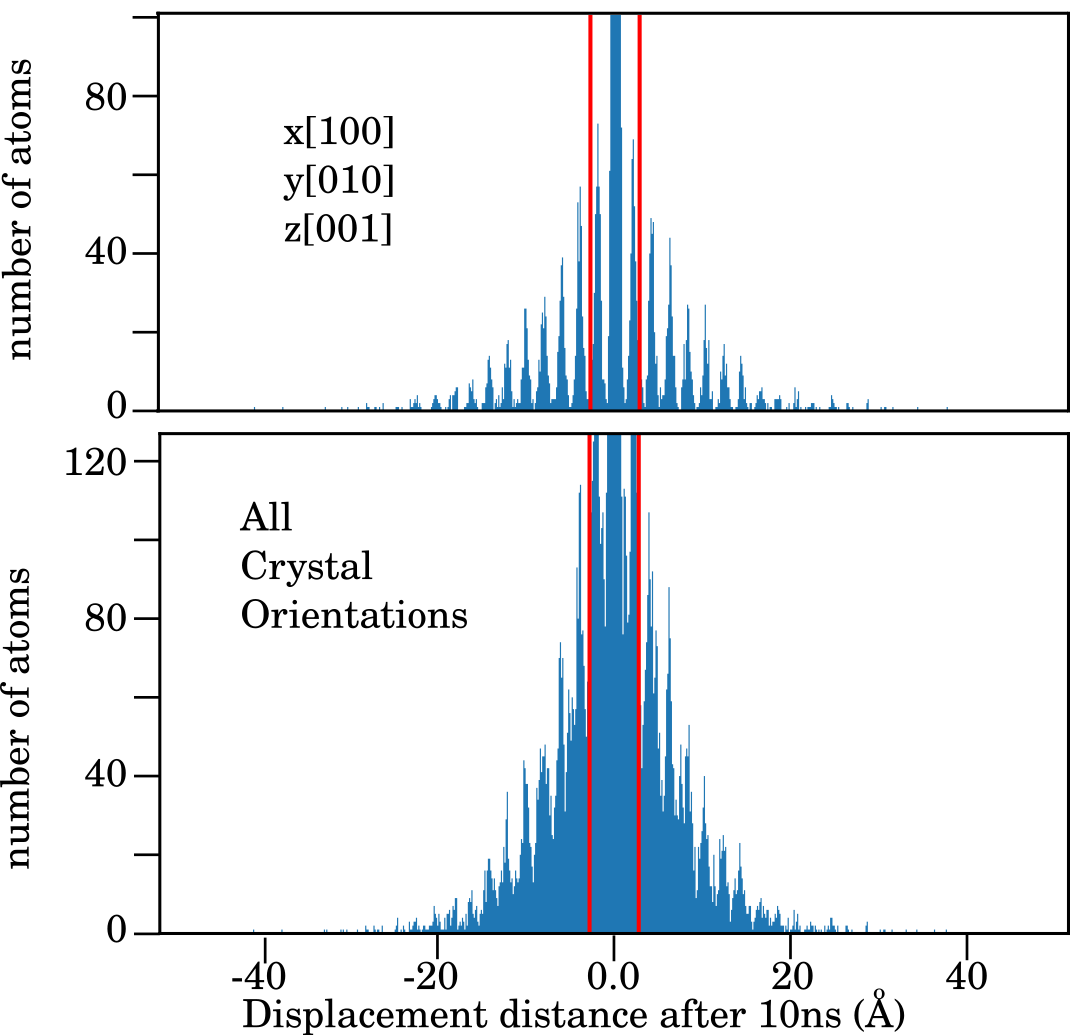}
    \caption{\label{fig:gaussian_smooth}
    Distributions of atomic displacements in the $z$ direction after $10\ \mathrm{ns}$ at $900\ \mathrm{K}$ for cylinders of radius $34\ \text{\AA}$;
    the vertical red lines indicate the interatomic distance.
    Displacements on a cylinder with [$001$] along the $z$ direction (top) generally occur at integer multiples of a lattice translation vector.
    Combining the displacement data from simulations using the four orientations shown in Fig.\ \ref{fig:orientations} reduces the strength of these peaks.}
\end{figure}

\begin{figure}[t]
    \centering
    \includegraphics[width=0.98\columnwidth]{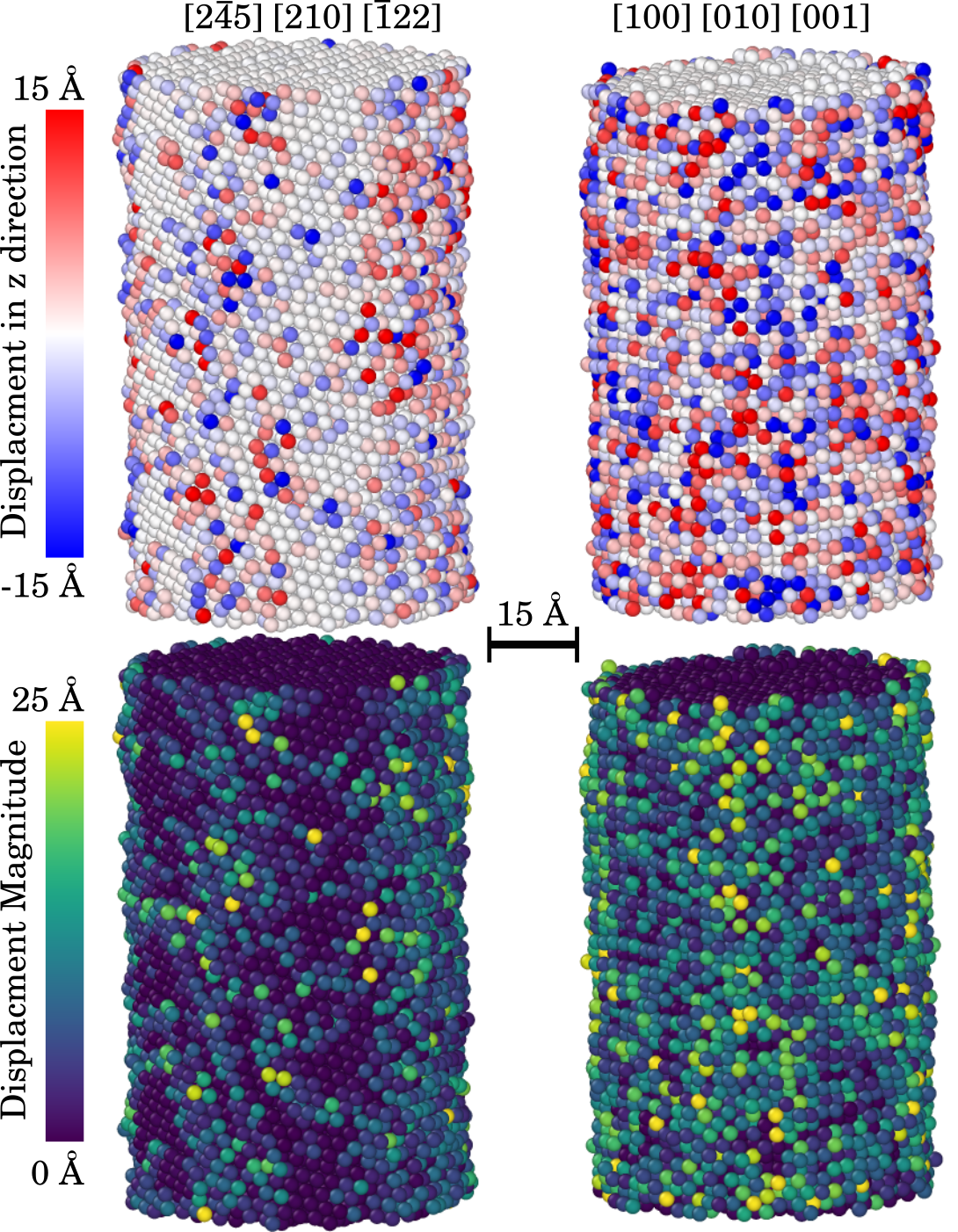}
    \caption{\label{fig:displ_vis}The $z$ components (top) and magnitudes (bottom) of atomic displacements on cylinders of radius $30\ \text{\AA}$ during a $10$ ns simulation at $900$ K.
    The displacement magnitude and the number of displaced atoms visibly depends on the crystallographic orientation (left and right).}
\end{figure}

Of course, there are limitations to this approach.
Equation \ref{equ:prob_distr_data} should only be used to analyze displacements along an eigendirection of the surface diffusivity tensor, since it is only under these circumstances that the overall displacement of a random walker can be decomposed into a product of distributions in the form of Eq.\ \ref{prob_distr}.
The surface diffusivity is expected to have eigendirections aligned with the directions of the principle curvatures, though this is a reasonable assumption only if the surface diffusivity is defined as an average over crystal orientations.
Observe that in the top of Fig.\ \ref{fig:gaussian_smooth}, the distribution of $z$-displacements has a series of peaks corresponding to discrete jumps between atomic sites along the [$001$] direction; the orientation of the crystal lattice generally has a strong effect on the distribution of atomic displacements in any particular MD simulation, emphasizing the need to average over multiple crystal orientations.
This is more directly visible in Fig.\ \ref{fig:displ_vis}, where the atomic displacements visibly depend on the crystallographic orientation (all other simulation properties are the same).
Averaging over crystal orientations as in the bottom of Fig.\ \ref{fig:gaussian_smooth} not only mitigates this effect, but makes clear that the probability distribution can reasonably be modeled as a sum of Gaussians, with a narrow Gaussian (the peak is above the vertical limit of the image) corresponding to the bulk population and a wider Gaussian corresponding to the surface population.
For emphasis, fitting this probability distribution to Eq.\ \ref{equ:prob_distr_data} allows $D_S$, $D_B$, and $n_S$ to be found without needing to provide an explicit atomic-level prescription for what constitutes a surface atom.

Notice that a finite mobile surface atom lifetime would cause the apparent values of $n_S$ and $D_S$ to depend on the simulation duration, with $n_S$ increasing (as atoms join and leave the mobile surface atom population) and $D_S$ decreasing (as atoms spend a decreasing fraction of the simulation in the mobile state) with increasing simulation duration.
Increasing the simulation size and decreasing the duration to measure the instantaneous values of $n_S$ and $D_S$ is not feasible since the diffusion length needs to be at least several interatomic spacings for the distribution of atomic displacements to be reasonably approximated as a Gaussian distribution.
That said, the time-averaged flux of surface atoms over the simulation duration is equal to the average of the instantaneous fluxes, and since the instantaneous fluxes should be constant by time translation symmetry, the average flux should not depend on the simulation duration.
The average flux is also proportional to the product of the average concentration and average velocity of the diffusing species, meaning that the product $n_S D_S$ is expected to be independent of simulation duration even when $n_S$ and $D_S$ individually are not.
This idea is explicitly tested in Sec.~\ref{subsec:diff_results} below, and supports the existence of a finite mobile surface atom lifetime \cite{antczak2007jump}.


Equation \ref{equ:prob_distr_data} was fit to the simulation data by maximizing the log likelihood function
\begin{equation}
\label{equ:loglikeli}
\ell(D_S, D_B, n_S) = \sum_{i}\log P(x_i, t | D_S, D_B, n_S)
\end{equation}
with respect to the parameters $D_S$, $D_B$, and $n_S$, where $x_i$ is the displacement of the $i$th atom in either the $\nvec{z}$ or $\nvec{\theta}$ direction after $10\ \mathrm{ns}$.
This equation can be explicitly differentiated with respect to these quantities, allowing the maximization to be performed using a modified Brent's method \cite{brent2013algorithms}.
The uncertainty in the parameter estimates was found using bootstrapping where an auxiliary data set is generated by sampling from the original data set with replacement and the parameters are estimated using the auxiliary data set.
This procedure is repeated a sufficient number of times to be able to characterize the mean and variance of the resulting distributions of estimated parameter values.
This was implemented using $80$ auxiliary data sets, each of the same size as the original data set, and the resulting distributions of $D_S$, $D_B$, and $n_S$ are reported in Sec.~\ref{sec:resultsanddisc} below.

\subsection{Activation Energy and Temperature}

The values of $D_S$ and $n_S$ were found by the procedure in Sec.\ \ref{sec:diffusion} for a variety of temperatures.
$D_S$ is modeled as having a standard Arrhenius dependence on temperature:
\begin{equation}
\label{equ:activate}
D_S = D_S^0\exp\left(-\frac{\Delta H_S}{k_BT}\right)
\end{equation}
where $D_S^0$ is the diffusion prefactor, $\Delta H_S$ is the activation enthalpy, $k_B$ is Boltzmann's constant, and $T$ is the absolute temperature.
Notice that $D_S^0$ depends on a distribution of surface atom configurations and migration events, and since these change with the surface mean curvature, $D_S^0$ could depend on the surface mean curvature as well.
The fraction of surface atoms $n_S$ does not have a standard form but is modeled by:
\begin{equation}
\label{equ:nsenergy}
n_S = \alpha_0 \frac{2r_aA}{V} \exp\left(-\frac{E_F}{k_BT}\right)
\end{equation}
where $E_F$ is the formation energy of a mobile surface atom, $A$ and $V$ are the total surface area and volume of the gold structure (e.g., cylinder), respectively, and $r_a$ is the radius of a gold atom.
The prefactor can be motivated by observing that $2r_a A/ V$ is roughly the fraction of atoms in the simulation cell that belong to a single surface monolayer.
This implies that $\alpha_0$ is a dimensionless factor proportional to the number of surface monolayers from which an atom could become a mobile surface atom.

The activation energies and prefactors in Eqs.\ \ref{equ:activate} and \ref{equ:nsenergy} can be found from the slopes and intercepts of semi-log plots of $D_S$ and $n_S$ as functions of temperature.
Simulations were performed for each combination of curvature and crystal orientation at four different temperatures, namely, $900$, $950$, $1000$, and $1050\ \mathrm{K}$.
This temperature range was selected for being the highest possible without initiating surface melting and changing the mass transport mechanism, the intention being to increase the number of atomic jumps observed within a fixed simulation duration of $10\ \mathrm{ns}$.
For comparison, coarsening of experimental np-Au samples generally occurs at temperatures of $500$-$700\ \mathrm{K}$ over times on the order of minutes to hours~\cite{Seker2007}.


\section{Results and Discussion}
\label{sec:resultsanddisc}


\subsection{Atomic Potentials}

As stated earlier, the surface energy of np-Au could substantially affect the mass transport mechanism.
This requires careful selection of the interatomic potential since one where the surface energy is unrealistically large could drive the generation of dislocations and mass transport by plastic deformation.
Alternatively, an interatomic potential where the binding energy of surface atoms is unrealistically small could increase the surface atom mobility as a consequence of premature surface melting.

\begin{figure}
    \begin{centering}
    \includegraphics[width=0.98\linewidth]{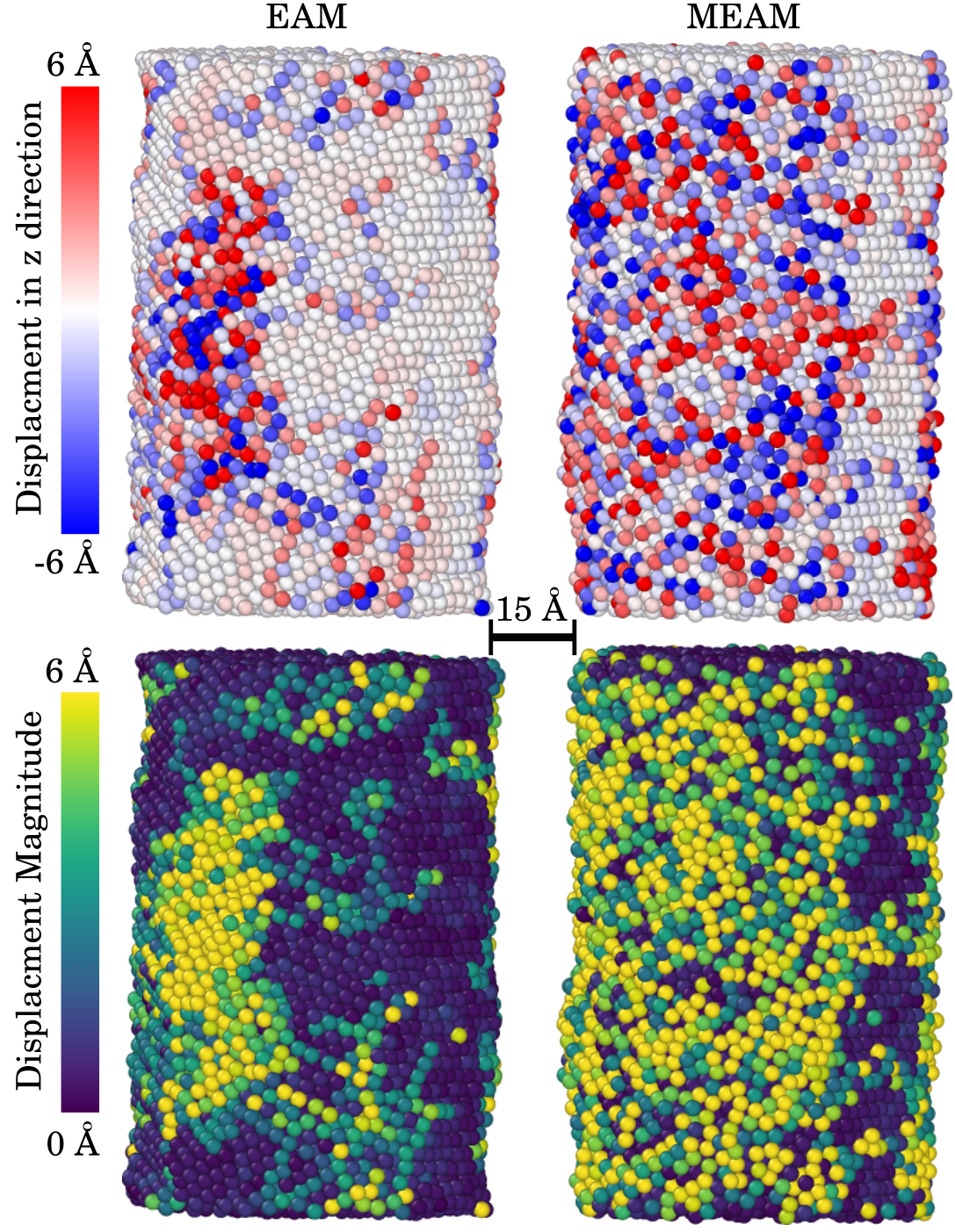}
    \caption{\label{fig:MEAM_EAM}Atomic displacements during $1$ ns as simulated using the EAM potential (left) or the MEAM potential (right).
    Both simulations started with identical atomic coordinates for a cylinder of radius $30\ \text{\AA}$ and a [$1\bar{7}4$] [$\bar{1}12$] [$311$] orientation and were relaxed for $10$ ns at $1000$ K.}
    \end{centering}
\end{figure}

\begin{figure}
    \begin{centering}
    \includegraphics[width=0.98\linewidth]{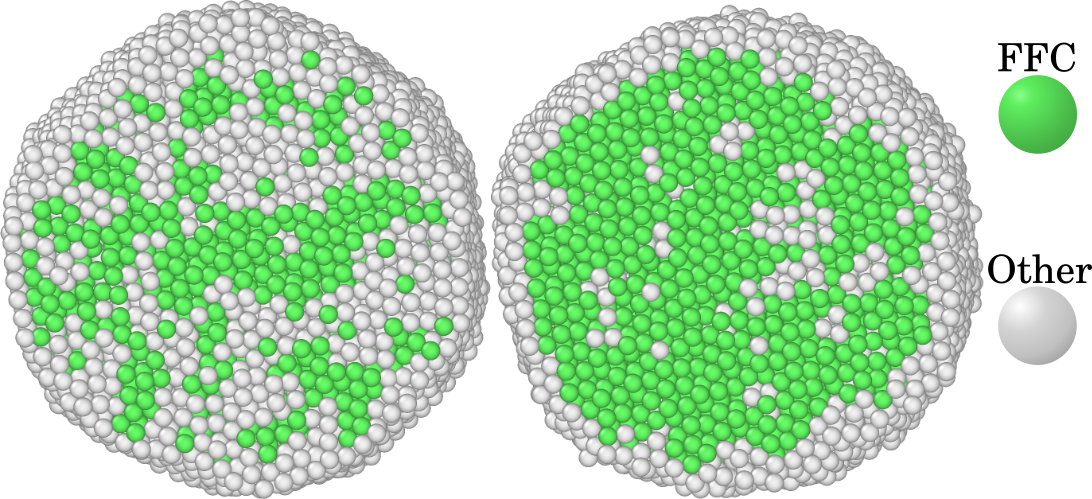}
    \caption{\label{fig:MEAM_EAM_depth} The Au cylinder on the left shows the atomic coordination for each atom after a $10$ ns simulation at $1000$ K using the EAM potential (left) or the MEAM potential (right).}
    \end{centering}
\end{figure}

Two otherwise identical simulations were performed to compare the effects of using an EAM potential \cite{Grochola2005} and a MEAM potential \cite{Lee2003} for gold.
Both simulations contained a pillar of $13,853$ gold atoms, with the pillar axis along the $z$ axis, orientated such that the crystallographic directions [$1\bar{7}4$], [$\bar{1}12$], and [$311$] were respectively aligned with the $x$, $y$, and $z$ axes of the simulation box.
Both simulations were thermostated at $1000$ K and run for a total of $11$ ns, with the atomic displacements over the final $1$ ns of the simulation plotted in Fig.~\ref{fig:MEAM_EAM}.
It is noteworthy that the atoms in the simulation using the EAM potential can be seen by the common neighbor analysis in Fig.~\ref{fig:MEAM_EAM_depth} to be significantly more disordered further from the surface compared to those in the simulation using the MEAM potential.
Along with the observation that atoms displace in the EAM simulation significantly deeper from the surface, this suggests that there could be surface melting at temperatures below the bulk melting point of gold in the EAM simulations.


\begin{figure}[t]
    \centering
    \includegraphics[width=0.98\columnwidth]{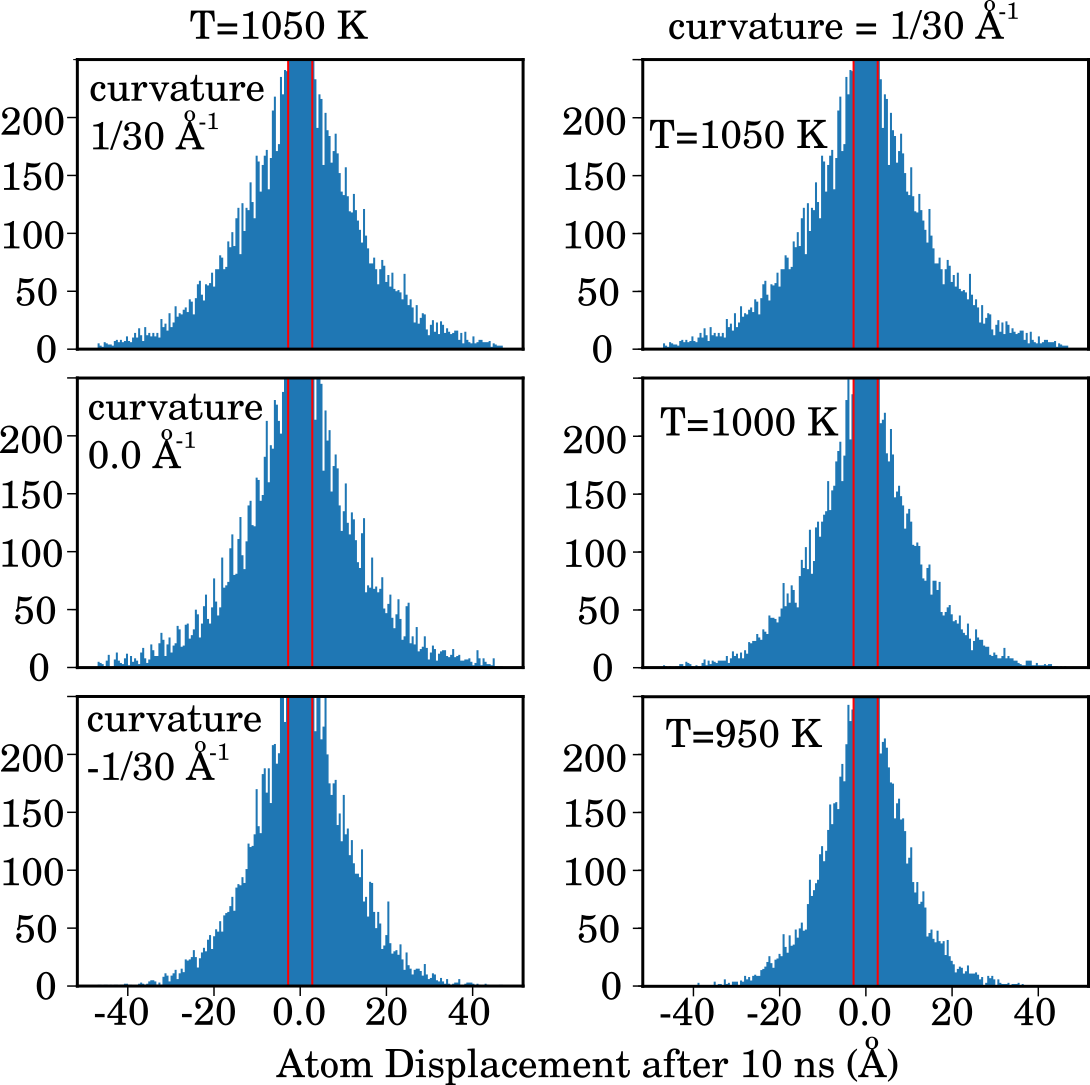}
    \caption{\label{fig:hist_disp}Histogram showing the distribution of atomic displacements in the $\nvec{\theta}$ direction after $t = 10$ ns, with one series of varying curvature (left) and one of varying temperature (right).
    The distribution of surface atom displacements measurably broadens with increasing curvature and temperature (up the figure).
    The figures have been cropped in the vertical direction since the number of negligible displacements of bulk atoms is two orders of magnitude higher.}
\end{figure}

The model we are using to extract diffusion coefficients from atomic displacements requires combining data from enough simulations with different crystallographic orientations to give a displacement distribution that approximates a smooth Gaussian. 
Figure \ref{fig:hist_disp} shows several such distributions, each including data from all four orientations in Fig.\ \ref{fig:orientations} simulated at the indicated temperature and curvature;
these four orientations appear to be sufficient to give a distribution that resembles a Gaussian, allowing Eq.\ \ref{equ:prob_distr_data} to be reliably fit to the data.
The histograms are cropped to better show the part of the distribution relevant to surface diffusion as there are orders of magnitude more counts for the negligible displacements of the bulk atoms.
Observe that the histograms measurably broaden with increasing temperature and curvature (up the figure).
Whereas the broadening with temperature is expected since atomic displacement is a thermally-activated event, the broadening with curvature is less so since the population and diffusivity of mobile surface atoms are usually assumed to be curvature independent.





\subsection{Diffusion and Activation Energy Results}
\label{subsec:diff_results}


The first part of Eq.~\ref{equ:prob_distr_data} corresponds to the distribution of surface atom displacements, with $D_S$ proportional to the variance and $n_S$ equal to the probability mass of the distribution.
Physically, $D_S$ is related to the expected distance that a mobile surface atom travels in a given time, whereas $n_S$ is related to the expected number of mobile surface atoms.

\begin{figure}[t]
    \centering
    \includegraphics[width=0.98\columnwidth]{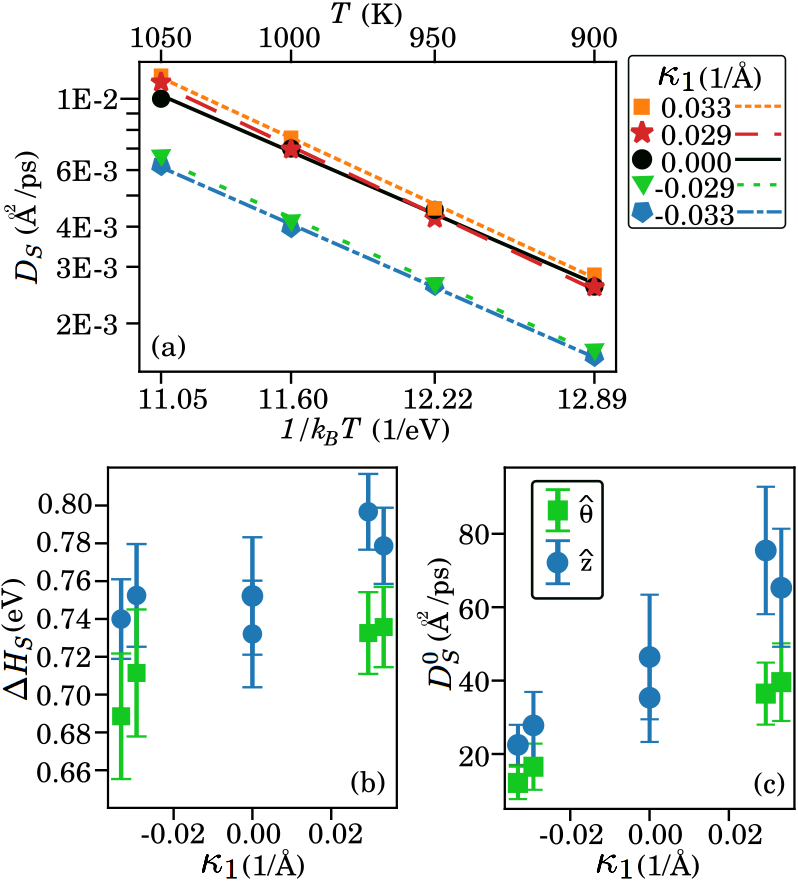}
    \caption{\label{fig:act_energy}(a) Semi-log plot of surface diffusion coefficients $D_S$ for several curvatures, found by fitting Eq.\ \ref{equ:prob_distr_data} to the displacements along $\nvec{z}$.
    With $D_S$ as in Eq.\ \ref{equ:activate}, (b) the activation energy of migration $\Delta H_S$ and (c) the surface diffusion coefficient prefactor $D_S^0$ are shown as functions of curvature $\kappa_1$ for displacements along $\nvec{z}$ and $\nvec{\theta}$.
    Error bars show the standard deviation of the bootstrapping results and are negligible for $D_S$.}
\end{figure}

\begin{table}
    \centering
    \resizebox{\columnwidth}{!}{%
    \begin{tabular}{llll}
    \hline
    Ref. & Method & \begin{tabular}[c]{@{}l@{}}Activation \\ Energy (eV)\end{tabular} & \begin{tabular}[c]{@{}l@{}}Prefactor \\ (\AA$^2$/ps)\end{tabular} \\ \hline
    ~\cite{Lin1989} & Profile Decay & 0.91 &  \\
    ~\cite{Mclean1968} & Scratch method & 2.82 & $8.0\times10^{10}$ \\
    ~\cite{Beszeda2004} & \begin{tabular}[c]{@{}l@{}}Thin-film agglom-\\ erations\end{tabular} & 0.64 & $3.0\times10^{-3}$ \\
    ~\cite{Gbel1995} & Scratch Method & 0.4 & $5.0\times10^{-4}$ \\
    ~\cite{Drechsler1981} & Crystalite profile & 0.35 & $5.0\times10^{-5}$ \\
    ~\cite{Porath1995} & Profile Decay & 1.1 & 5  to 10 \\
    ~\cite{Liu1991} & MD simulation &  &  \\
     & (100) facet & 0.64, 0.84 & 8, 35 \\
     & (111) facet & 0.021, 0.038 & 2, 3.7 \\
     & (311) facet & 0.35, 0.42 & 8, 36 \\
     & (331) facet & 0.26, 0.34 & 6, 9.6 \\
    ~\cite{Baletto2000} & MD simulation &  &  \\
     & exchange & 0.15, 0.24 &  \\
     & jump & 0.11, 0.29, 0.34 &  \\
     & (100) facet & 0.41 &  \\
    ~\cite{Boisvert1993} & MD simulation &  &  \\
     & (111) facet & 0.013, 0.014 & 0.45, 0.9 \\ \hline
\end{tabular}%
}
\caption{A broad range of activation energies for surface diffusion have been reported in the literature, though this could indicate that they are measuring different populations of events~\cite{Liu1991}. 
The lower activation energies mostly correspond to single adatom diffusion events.}
\label{tab:reference_Ds}
\end{table}

The surface diffusion coefficients plotted in Fig.~\ref{fig:act_energy}(a) were found by the procedure described in Sec.\ \ref{sec:diffusion}, and depend on the assumption that the sets of atoms belonging to the bulk and surface populations are fixed.
Since $D_S$ is expected to follow the Arrhenius temperature dependence in Eq.~\ref{equ:activate}, the values are plotted with respect to $1/(k_B T)$ for each $\kappa_1$.
The activation energy for surface diffusion $\Delta H_S$ and the corresponding prefactor $D_S^0$ were found from the slopes and the intercepts of the plots of $D_S$ and are shown as functions of curvature in Figs.~\ref{fig:act_energy}(b) and (c).
These reveal that $\Delta H_S$ appears to have a weak dependence on $\kappa_1$, with the magnitudes of the slopes in Fig.~\ref{fig:act_energy}(a) slightly increasing with curvature.
However, the apparent dependence on curvature is likely a consequence of a finite expected lifetime of a mobile surface atom, with the true surface diffusivity being independent of curvature as is discussed in more detail below.
The $\Delta H_S$ values found range from $0.69$ to $0.79$ eV and are close to the median of the previously-reported experimental values in Table~\ref{tab:reference_Ds}. 
More specifically, the experimental values range from $0.35$ to $2.82$ eV depending on the measurement method, and the values from MD simulations range from $0.013$ to $0.84$ depending on the specific migration event considered.
It is significant that the activation energy for any particular migration event in an MD simulation does not adequately reflect the spectrum of events that occur in either an experimental system or our MD simulations.
Indeed, one benefit of the method developed in Sec.\ \ref{sec:diffusion} to measure surface diffusion is that the resulting diffusivities naturally account for this spectrum of events.

The surface diffusion prefactor $D_S^0$ shown in Fig.~\ref{fig:act_energy}(c) appears to be an increasing function of curvature, but the dependence is small enough that it would be difficult to measure experimentally.
$D_S$ in Fig.~\ref{fig:act_energy}(a) does shift upward with increasing $\kappa_1$, though the trend is irregular with the curves for $\kappa_1 > 0.0$ and $\kappa_1 = 0.0$ separated from those for $\kappa_1 < 0.0$.
While this suggests that $D_S^0$ (the vertical intercept of the $D_S$ curves) could increase with curvature, the region plotted in Fig.~\ref{fig:act_energy}(a) is far enough from the vertical axis that the intercept value depends more on the slope than the vertical offset.
That is, the apparent value of $D_S^0$ is highly sensitive to changes in $\Delta H_S$, and a small systematic error in $\Delta H_S$ would be more than enough to account for the apparent curvature dependence of $D_S^0$ in Fig.~\ref{fig:act_energy}(c).
The same sensitivity occurs for experimental diffusion measurements as well, with the surface diffusion prefactors in Table~\ref{tab:reference_Ds} increasing by 15 orders of magnitude as the activation energies increase from $0.35$ to $2.82$ eV.

\begin{figure}[t]
    \centering
    \includegraphics[width=0.98\columnwidth]{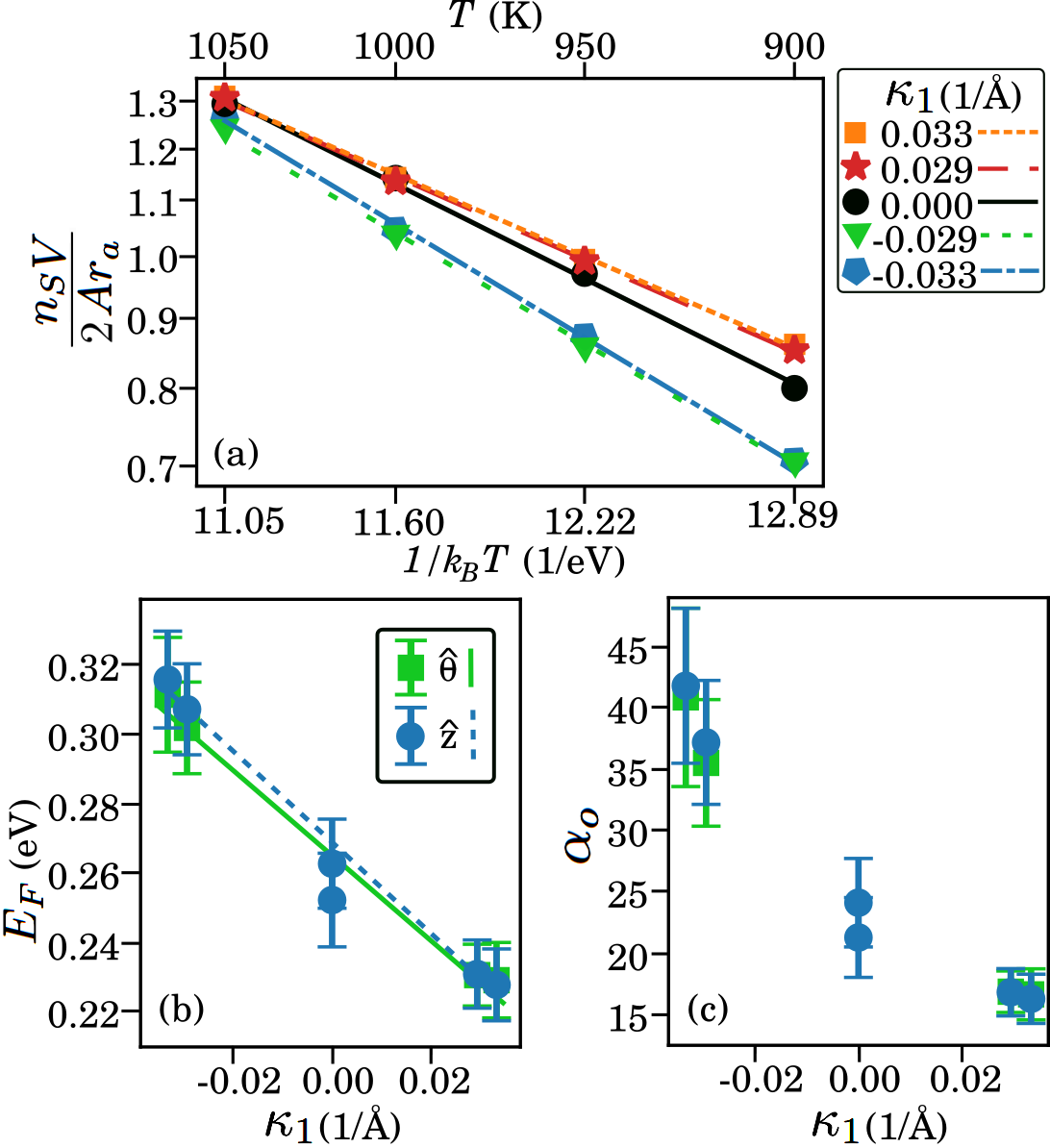}
    \caption{\label{fig:ns_act_energy}Semi-log plot of $n_S$ scaled by $V / (2 A r_a)$ for several curvatures (top), found by fitting Eq.\ \ref{equ:prob_distr_data} to the displacements along $\nvec{z}$.
    With $n_S$ as in Eq.\ \ref{equ:nsenergy}, the formation energy of a mobile atom $E_F$ (bottom left) and the prefactor $\alpha_0$ (bottom right) are shown as functions of curvature for displacements along $\nvec{z}$ and $\nvec{\theta}$.
    The linear fits for $E_F$ as a function of curvature are provided as guides for the eye.
    Error bars show the standard deviation of the bootstrapping results and are negligible for $n_S$.}
\end{figure}

\begin{figure}[t]
    \centering
    \includegraphics[width=0.98\columnwidth]{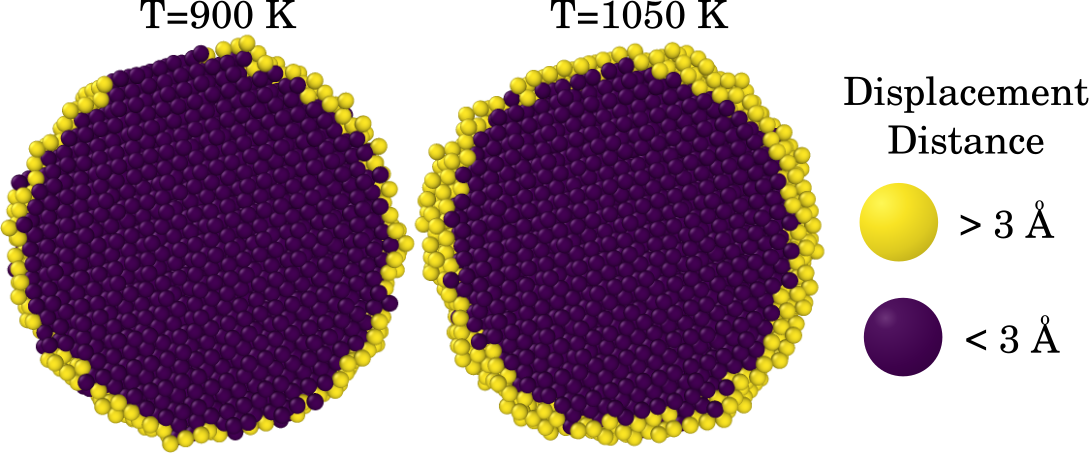}
    \caption{\label{fig:num_surf_atom}Cross-sectional view of a single Au cylinder with a radius of $34$ \AA\ simulated at $900$ K (left) and $1050$ K (right).
    Atoms that have displaced more than $3$ \AA\ after a $10$ ns simulation are colored in yellow.
    For reference, $2.8$ \AA\ is the approximate diameter of a gold atom.}
\end{figure}

As described above, $n_S$ is the other significant quantity derived by fitting atomic displacement data to Eq.~\ref{equ:prob_distr_data}, and is precisely defined as the orientation-averaged fraction of atoms whose displacements along one of the surface's principal directions are characterized by the diffusivity $D_S$;
these atoms will be referred to as mobile surface atoms for brevity.
This fraction depends on the total number of atoms in the simulation though, which for the punch hole geometry in particular is arbitrary.
Hence, with $n_S$ modeled as in Eq.~\ref{equ:nsenergy}, $n_S V / (2 r_a A) = \alpha_0 \exp[-E_F / (k_B T)]$ is plotted as a dimensionless quantity that roughly indicates the number of effective mono-layers of mobile surface atoms.  
Note that the values for $n_S V / (2 r_a A)$ plotted in Fig.~\ref{fig:ns_act_energy}(a) are all of order one, consistent with the expectation that roughly a single mono-layer of surface atoms are able to diffuse by a short-circuit pathway.
That said, the number of mobile surface atoms can exceed the number of atoms in a single surface mono-layer at higher temperatures;
Figure \ref{fig:num_surf_atom} shows that one effective mono-layer of displaced atoms at $900$ K increases to nearly two mono-layers at $1050$ K for this particular crystallographic orientation.
Notice that this does not necessarily indicate that the concentration of mobile surface atoms is higher at $1050$ K than at $900$ K at every instant in time, but rather that the population of atoms that displaces on the surface at some point during the simulation (and therefore contributes to $n_S$) increases with temperature.
Indeed, the possibility of a finite mobile surface atom lifetime is supported by the literature \cite{antczak2007jump} and by direct observations of atomic exchange between the two outermost effective mono-layers in the simulation at $1050$ K shown in Fig.~\ref{fig:num_surf_atom}.
This reinforces that $n_S V / (2 r_a A)$ should be considered as indicating the total number of atoms that belonged to the mobile surface atom population at any point during the simulation, rather than the surface concentration of atoms that are mobile at every instant in time.
The distinction is subtle but critical to an accurate interpretation of our results since, for a fixed surface displacement, the true diffusivity of a surface atom increases as the time during which that atom belongs to the mobile surface population decreases.

The curves for $n_S V / (2 r_a A)$ in Fig.~\ref{fig:ns_act_energy}(a) shift upward and decrease in slope with increasing $\kappa_1$,
appearing to all converge to a single value that is independent of curvature at a temperature higher than $1050$ K but much lower than the bulk melting point of gold at $1337$ K ($1440$ K for the MEAM potential~\cite{Lee2003}).
This would be quite surprising since the effect of curvature on the mobile surface atom population would then be reversed at still higher temperatures, with the apparent population of mobile surface atoms being larger on negatively-curved surfaces than positively-curved ones near the melting point.
One possible resolution is that there could be surface melting at a temperature below that of the bulk material \cite{guenther2014models}, causing the point of convergence to coincide with a change of transport mechanism and making the apparent population of mobile surface atoms larger on positively-curved solid surfaces at all temperatures.
It is also significant that the dependence of the slope on $\kappa_1$ makes the mobile surface atom populations on positively-curved and negatively-curved surfaces increasingly diverge at lower temperatures.
Extrapolating the curves to $600$ K, the temperature that is often used in thermal annealing experiments, suggests that the apparent population of mobile surface atoms could be nearly twice as high on a convex surface as on a concave one.

The slopes of the curves for $n_S V / (2 r_a A)$ in Fig.~\ref{fig:ns_act_energy}(a) are precisely the energy $E_F$ in Eq.~\ref{equ:nsenergy}, and are plotted with respect to $\kappa_1$ in Fig.~\ref{fig:ns_act_energy}(b).
$E_F$ is interpreted as the characteristic energy required for an immobile surface atom to become mobile.
While we do not speculate as to the mechanism, the magnitude of $E_F$ ($\about 0.27$ eV) is comparable to the single bond energies of $0.15$ to $0.30$ eV routinely used in KMC simulations~\cite{Erlebacher2011,Li2019}.
Moreover, Fig.~\ref{fig:ns_act_energy}(b) shows that $E_F$ is a clearly decreasing function of curvature given the magnitude of the random error.
This is consistent with the interpretation of $E_F$ above since the number of nearest neighbors of a surface atom on average decreases with increasing curvature, reducing the energetic barrier for an atom to join the population of mobile surface atoms with increasing curvature.
Notice that for the instantaneous concentration of mobile surface atoms to be constant, the increased rate at which atoms join the population of mobile surface atoms with increasing curvature would require each atom to belong to that population for a shorter time, meaning that the lifetime of a mobile surface atom would need to decrease with increasing curvature.


A reasonable question is why identical reasoning should not apply to the activation energy for surface diffusion $\Delta H_S$;
if a decrease in $E_F$ with curvature reflects a decrease in the average coordination number of a surface atom, then one could expect $\Delta H_S$ to decrease with curvature since the binding energy of the atoms to the surface would decrease as well.
One possible explanation relates to the proposed finite lifetime of a mobile surface atom.
If the rate of exchange of atoms between the bulk and mobile surface atom populations increases with curvature, then the lifetime of a mobile surface atom (i.e., the period during which the atom is able to diffuse on the surface) would decrease with curvature.
This could reduce the apparent diffusivity that would be calculated given the assumption that the set of atoms belonging to the mobile surface atom population is fixed, with the result that the apparent activation energy of surface diffusion increases as the mobile surface atom lifetime decreases.
It is significant that this could occur despite the average coordination number of a surface atom decreasing with increasing curvature depending on the relative strengths of the effects.

The quantity plotted in Fig.~\ref{fig:ns_act_energy}(c) is $\alpha_0$, corresponding to a reference number of effective surface mono-layers in Eq.\ \ref{equ:nsenergy}.
The only physical significance ascribed to this quantity is that $\alpha_0 \exp[-E_F / (k_B T)]$ should equal the number of effective mono-layers of mobile surface atoms at the temperature at which the conjectured surface melting occurs.
While $\alpha_0$ does appear to be a decreasing function of curvature, the dependence is not believed to be significant, with a small systematic error in $E_F$ being more than enough to account for the curvature dependence of $\alpha_0$ in Fig.~\ref{fig:ns_act_energy}(c).


It is worthwhile commenting briefly on the two values plotted for the $z$ direction in Figs.~\ref{fig:act_energy}(b), \ref{fig:act_energy}(c), \ref{fig:ns_act_energy}(b) and \ref{fig:ns_act_energy}(c) for $\kappa_1 = 0$.
There are no results for the $\theta$ direction since this direction is poorly defined in the slab geometry.
Instead, the $\nvec{x}$ displacements were aggregated into a separate data set, with the derived values colored the same way as for $\nvec{z}$ displacements since the results along the $z$ and $x$ directions should coincide given sufficient sampling of crystal orientations.
As for the simulations of cylinders and punches, the difference between diffusion along the $\theta$ and $z$ directions (corresponding to $\hat{\kappa}_1$ and $\hat{\kappa}_2$ respectively) is subtle. 
It appears in Figure~\ref{fig:act_energy}(b) that $\Delta H_S$ is slightly higher for diffusion in the $z$ direction;
this is attributed to the relatively small number of crystal orientations sampled, since the magnitude of the difference is comparable to that between the $\nvec{x}$ and $\nvec{z}$ displacements for the slab geometry.

\begin{figure}[t]
    \centering
    \includegraphics[width=0.98\columnwidth]{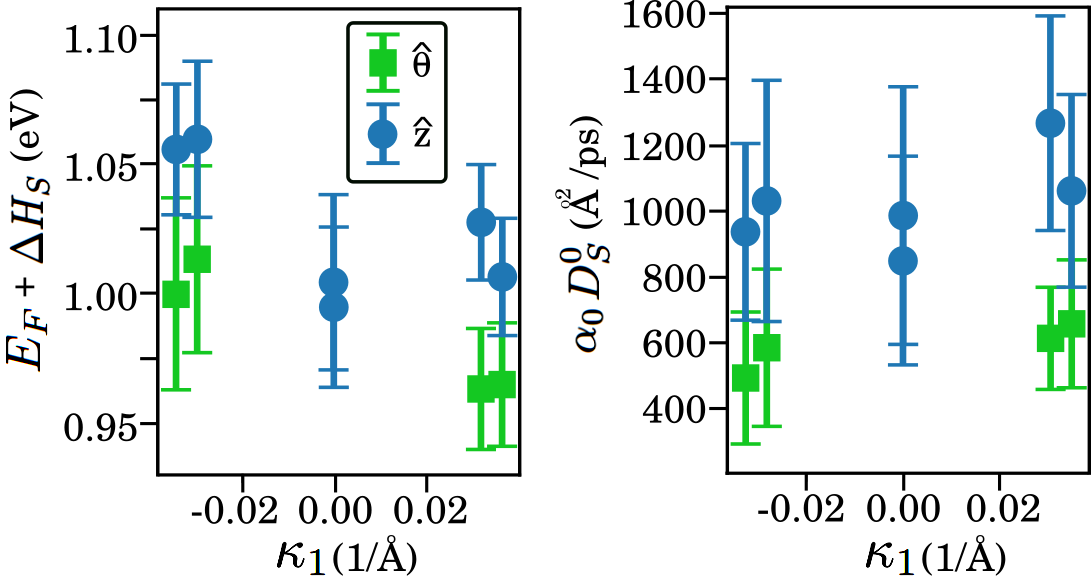}
    \caption{\label{fig:mass_trans_energy} The rate of mass transport by surface diffusion is expected to have an Arrhenius dependence with an activation energy $E_F + \Delta H_S$ and a prefactor proportional to $\alpha_0 D^0_S$.
    These quantities are independent of curvature to within the uncertainty estimates, despite the apparent curvature dependencies in Figs.\ \ref{fig:act_energy} and \ref{fig:ns_act_energy}.}
\end{figure}

Transport theory indicates that the rate of mass transport by surface diffusion should be directly proportional to the product of the number of mobile surface atoms and the surface diffusivity.
Given the Arrhenius dependencies of Eqs.\ \ref{equ:activate} and \ref{equ:nsenergy}, this implies that the rate of mass transport by surface diffusion should have an Arrhenius dependence with an activation energy of $E_F + \Delta H_S$ and a prefactor proportional to $\alpha_0 D^0_S$.
These quantities are plotted in Fig.\ \ref{fig:mass_trans_energy} as functions of the curvature, and appear to be independent of curvature within the estimated uncertainty.
While it is initially surprising that the curvature dependencies of $D_S$ and $n_S$ should be nearly equal and opposite, decades of literature has assumed that mass transport by surface diffusion is independent of curvature \cite{Herring1950}.
This means that the real question is why the analysis performed here found $D_S$ and $n_S$ to have any curvature dependence at all.

\begin{figure*}[t]
    \centering
    \includegraphics[width=1.54\columnwidth]{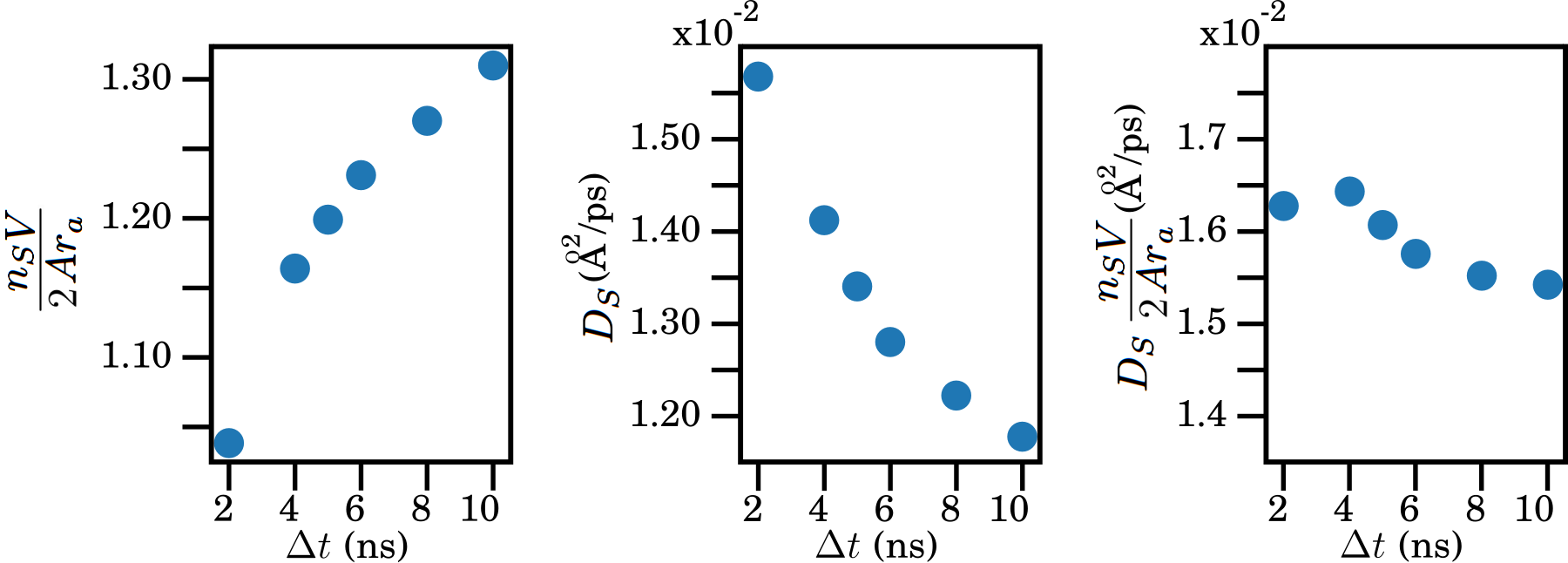}
    \caption{\label{fig:time_dependence}The values plotted are calculated from a data set made up of atomic displacements found after time $\Delta t$. The data set sampled from for each $\Delta t$ is made up from four different simulations of a 30 \AA~cylinder at $1050$ $K$ each with different crystal orientations. Mobile atom exchange with an immobile atom population increases with time and this is reflected in the calculated values for $D_S$ and $n_S$ though their product is relatively constant.}
\end{figure*}

As suggested in Sec.~\ref{sec:diffusion}, we propose that the apparent dependence on curvature is not introduced by the simulation setup, but by one of the assumptions behind Eq.\ \ref{equ:prob_distr_data}.
Specifically, Eq.\ \ref{equ:prob_distr_data} assumes that the set of atoms belonging to the mobile surface atom population is fixed for the entire simulation duration, and that no atoms are exchanged with the immobile bulk atom population.
If this were true, then the variance of the displacement distribution of the fixed population of mobile surface atoms would indeed increase as $t D_S$ as is assumed above.
However, if only the concentration of mobile surface atoms is fixed, then individual atoms could be exchanged between the mobile surface atom and immobile bulk populations at a curvature-dependent rate over the simulation duration.
Increasing this rate would decrease the mobile surface atom lifetime and, as the simulation duration increases relative to this lifetime, have two significant effects on derived quantities.
First, the apparent population of mobile surface atoms would increase since many atoms would join and leave the population of mobile surface atoms at some point during the simulation.
Second, the apparent surface diffusivity would decrease since any single mobile surface atom would displace on the surface for a decreasing fraction of the simulation.
Figure~\ref{fig:time_dependence} shows that this appears to be the case, with $n_S V / (2 r_a A)$ on the left increasing and $D_S$ in the middle decreasing with the simulation duration $\Delta t$.
The reasoning in Sec.~\ref{sec:diffusion} further suggests that the flux of surface atoms and the product $n_S D_S$ should be independent of $\Delta t$, and this is also supported by the plot on the right of Fig.~\ref{fig:time_dependence} after an initial transient related to the characteristic time for an atomic jump.
Notice that $n_S$ and $D_S$ are expected to approach asymptotic values for simulation durations much longer than the mobile surface atom lifetime when individual atoms can join and leave the mobile surface atom population multiple times, but a direct investigation is not computationally feasible at present.

The apparent curvature dependence of $D_S$ and $n_S$ can therefore be explained as a consequence of an increasing rate of exchange between the mobile surface atom and immobile bulk atom populations with increasing curvature.
Notice that the overall rate of mass transport by surface diffusion does not need to depend on this rate of exchange;
the instantaneous concentration of mobile surface atoms and the rate at which they jump to neighboring surface sites could easily be independent of both the rate of exchange and the surface curvature.
Unfortunately, there does not seem to be any way to directly measure these quantities by means of MD simulations, at least not without explicitly defining what is meant by a mobile surface atom, and (as already described in Sec.\ \ref{sec:diffusion}) it is not obvious whether this can be done in an unbiased way.
Our conclusion is that this unexpected phenomenon could make the apparent surface diffusivity reported by simulation studies dependant on not only the surface curvature but on the simulation time (relative to the mobile surface atom lifetime).
Moreover, making the standard assumption that the apparent population of mobile surface atoms is constant and independent of curvature and temperature could result in the activation energy for mass transport by surface diffusion being reported as $\Delta H_S$ alone, differing by $E_F$ ($\about 0.26$ eV) from the value reported here.

\section{Conclusion}
\label{sec:conclusion}

Many of the desirable properties of nanoporous metals in general, and np-Au in particular, are a consequence of an exceptionally high specific surface area $S_A$.
This high $S_A$ provides the driving force for the coarsening of the microstructure at moderate temperatures by surface diffusion, though directly measuring mass transport rates along individual ligaments by in-situ imaging is challenging due to the small features and short times involved.
An alternative approach is to infer the surface diffusivity by means of a scaling argument, assuming that the microstructure coarsens in a self-similar way.
However, there is no consensus in the literature that the microstructure coarsening is actually self-similar, and the activation energies found by this procedure can be as low as $0.353$ eV~\cite{Jeon2017} or as high as $1.33$ eV~\cite{McCue2018}.
This results in an uncertainty in the rate of mass transport by surface diffusion that makes it difficult to precisely predict the microstructural change that would result from a given thermal annealing schedule, or more generally, to construct processing routes to a nanoporous material with a desired microstructure.

Given this situation, atomistic simulations are a natural candidate to measure the rates of mass transport by surface diffusion in a simulated material.
Connecting such simulation results to the quantities that appear in classical transport theory is not trivial though, with the definition of various quantities like the chemical potential remaining unclear at the atomic scale.
The approach developed here compares the distribution of atomic displacements in the absence of a chemical potential gradient to that expected for an anisotropic random walker.
The resulting values for the surface diffusivity and mobile surface atom population appear to suggest subtle dependencies on the surface mean curvature, despite the overall rate of mass transport by surface diffusion being curvature independent.
It is proposed that the apparent curvature dependence of the surface diffusivity and mobile surface atom population is a consequence of the surface atoms having a finite lifetime during which they remain mobile (a possibility supported by the literature \cite{antczak2007jump}), after which they are exchanged with one of the population of effectively immobile bulk atoms.
This could occur in such a way that the concentration of mobile surface atoms remains constant, making the overall rate of mass transport by surface diffusion independent of the rate of exchange.
While an increasing rate of exchange with increasing surface curvatures would be able to explain our results, directly quantifying the rate of exchange would require a precise definition of what constitutes a mobile surface atom, and is not performed here.

The rate of mass transport is directly proportional to the product of the surface diffusivity and the population of mobile surface atoms, and is found in Fig.\ \ref{fig:mass_trans_energy} to have an activation energy of $\about 1.01$ eV, close to the median of the experimentally-determined activation energies reported in Table \ref{tab:reference_Ds}.
This includes an activation energy of $\about 0.27$ eV related to the activation energy for the exchange mechanism and the apparent dependence of the mobile surface atom population on temperature.
Given that the concentration of mobile surface atoms is conjectured to be constant, it would be easy for a researcher studying surface diffusion with MD simulations to assume that the population of mobile surface atoms should also be constant, and that the rate of mass transport should depend on temperature only through the surface diffusivity.
Measuring the surface diffusivity by comparing the displacement distribution of surface atoms to that of an anisotropic random walker would then lead to the the activation energy for mass transport by surface diffusion being incorrectly reported as only $\about 0.74$ eV, a difference that would significantly change the predicted thermal annealing schedules required to achieve a particular ligament dimension in experiments.

Finally, it is significant that the activation energy for mass transport found here is appreciably higher than that for surface diffusion for any of the MD simulations reported in Table \ref{tab:reference_Ds}.
Overall, this study underlines the importance of considering the possible effects of a range of migration events and the finite lifetime of a mobile surface atom when extracting surface diffusion values in simulated materials.

\section*{Data Availability}

The raw and processed data required to reproduce these findings are available for download in the supplementary materials.

\section*{Acknowledgements}

C.M.E, S.M.S., E.S and J.K.M.\ were supported by the National Science Foundation under Grant No.\ 2003849.

\appendix


\end{document}